\documentclass[a4paper,100pt]{article}

\usepackage{authblk}
\usepackage[T1]{fontenc}
\usepackage[utf8]{inputenc}
\usepackage{graphicx}
\title{Scaling Law for Discharges in Z pinch Devices} 

\author{
    Miguel Cárdenas, Alejandro Nettle, and Leandro Nú\~{n}ez
}

\affil{
    Laboratorio de Investigación de Cómputo de Física \\
    Facultad de Ciencias Naturales y Exactas \\
    Universidad de Playa Ancha, Av. Leopoldo Carvallo 270, Valpara\'iso, Chile \\
    Email: \texttt{miguel.cardenas@upla.cl}
}
\date{} 

\begin{document}

\maketitle

\large
\maketitle
\begin{abstract}
We consider the snowplow model for studying discharges in Z pinch devices. In this context, to obtain a complete picture on the physics of those discharges, we replenish the term disregarded in the original formulation of the snowplow equations. This is, we now consider not  only the magnetic force inward on the current sheath but we take into account also a force outward on it. Such a force results from the kinetic pressure of the gas occluded by the current sheath.

The internal energy gained by the gas towards the end of the discharge depends on the full history of the dynamical variables of the system. In other terms, the internal energy is a functional of the dynamical variables of the system. We write down the expression for evaluating that internal energy and we present also the formula for computing the temperature of the gas.

On this basis, we derive a scaling law that relates the temperature the gas in a Z pinch experiment would attain with the energetic and spatial wingspan of the corresponding experimental setup.

\end{abstract}

\section{Introduction}
Fusion reactions take place when the kinetic energy of colliding atomic nuclei is high enough as to overcome their corresponding Coulomb barriers. Thermonuclear fusion results from heating a confined atomic gas to very high temperatures, so that that gas becomes, at first instance, totally ionized. Of course, for setting a gas in a condition to attain a number of fusion reactions it is not enough to transform that gas  into a plasma. Such a plasma must be also kept sufficiently hot in order that its constituent particles are able to react \cite{teller}.

In principle, to impart energy to a plasma by neutron bombardment, radio waves and so on would not be a problem, however to impart energy to the plasma and to confine it altogether is what has serious limitations \cite{teller}.
In an attempt to overcome that difficulty, a number of apparatuses based on magnetic confinement have been devised 
\cite{bishop, hagler}.
In practice, an amount of neutral gas -contained within a chamber- is exposed to a very high voltage. This results into plasma breakdown followed by a large current that crosses the plasma. That column of current produces a strong magnetic field around itself. The coupling between the column of current and its own magnetic field results in a force inward that acts on the column of current and tends to contract it as a whole. In other words, the magnetic pressure set in by the magnetic field surrounding the column of current plays the role to compress the latter \cite{bennett, tonks, jackson}. 

After that, it is expected that at least a part of the energy passed from the voltage source to the gas transforms into internal energy of the latter. In this manner, the temperature of the gas would rise up and thereafter a number of fusion reactions would go on. 
Nevertheless, in the process of heating the plasma there are a number of unavoidable energy losses. For instance, as the particles of the plasma
moves chaotically, many electron-ion collisions take place. Those collisions imply severe deceleration of the electrons, so that an amount of the energy of the plasma is lost in the form of bremsstrahlung radiation. Those losses increase with temperature, because collisions are more frequent at higher temperatures and  they also increase when the plasma is contaminated by impurities \cite{spitzer}.   

For the case of a half and half deuterium-tritium mixture has been estimated that for the released energy of fusion exceeds the losses due to bremsstrahlung radiation, the temperature of the mixture has to be $4\,  keV$ or higher. Instead, for a system of pure deuterium the corresponding temperature should be of the order of $40\, keV$ \cite{glasstone}. Hereafter, we will express temperature $T$
in units of energy. This is, we will quote the value of $k_BT$ in $eV$ instead of quoting the value of $T$ in $K$.

Now then, to really sustain a thermonuclear reaction it is not sufficient to heat a plasma to a temperature as to compensate for bremsstrahlung radiation losses. In practice, that plasma must be heated much more until the numbers prescribed by the Lawson criterion are met \cite{lawson}. For instance, to sustain a half and half deuterium-tritium reaction the temperature of the mixture must be of the order of $25\,keV$ which goes far beyond the $4 \,keV$ necessary for overcoming radiation losses. In the case of deuterium-deuterium fusion, the corresponding temperature is an order of magnitude greater
than that \cite{glasstone}.
 
Hitherto, what some experiments -based on magnetic confinement methods- have shown is the onset of flashes of copious neutron emissions sometime during the discharges. However, those emissions do not come from genuine thermonuclear reactions. On the contrary, they result from collisions between some accelerated ions of the gas and the target formed by the relatively cold fraction of the gas \cite{anderson, hai, mc call}. In this respect, it is worth to mention that at present there is no a definitive straighforward experimental method for measuring accurately the temperature of the gas (\textit{i.e.} the plasma) in this kind of experiments. Therefore, to improve this situation in some way or another is an attending task.
Let us close this item by adding that up to now no magnetic confinement experiment has undeniably succeded to heat its working gas to temperatures within the range of thermonuclear interest.

A related issue concerns the possibility to find out a scaling law, if any, that connects the energetic and spatial size of a given experimental setup with the maximum temperature the working gas may eventually reach in the corresponding experiment. 

In the experiments we will be referring to, a bank of capacitors of energy $E_0$ is discharged into a plasma made up of $N$ constituent particles comprising ions and electrons. However, during a discharge it is only a fraction of the energy stored in the bank of capacitors that eventually transforms into internal energy of the plasma $U$. Furthermore, the quantity that determine essentially whether thermonuclear fusion takes place or not is the  temperature of the plasma that is given by the expression $k_BT=2/3(U/N)$. Hence, an experiment devised for the goal of approaching temperatures of interest for thermonuclear fusion, say $k_BT_{fusion}$, must be designed necessarily taking into account the constraint $2/3(E_0/N)>>k_BT_{fusion}$ otherwise there is no any chance to attain, by means of that experiment, thermonuclear fusion. 
 
Clearly, the task to fit the conditions for heating a system to temperatures within the useful thermonuclear range represents a great theoretical and empirical challenge. As a matter of fact, the several decades of research work on magnetic confinement fusion and also on inertial confinement fusion have not lead to the desired result. This is, to induce through those methods a thermonuclear fusion reaction similar to those reactions that go on in stars like our sun. Of course, this situation should not be a surprise because the thermonuclear reactions that take place in the sun occur in a fashion that does not quite match the dynamics observed in either magnetic confinement experiments
or inertial confinement experiments.    

Many experiments on discharges through gases have been assayed and subsequently they have been progressively optimized by trial and error essentially. The dynamics and electrical data gathered from the optimized experiments performed with either Z pinch devices \cite{post, kolb, ryutov} or plasma focus devices \cite{mather} (\textit{i.e.} a variant of the Z pinch apparatuses) have been reasonably well reproduced by means of a simple dynamical model known as snowplow model \cite{rosenbluth}. However, this model, owing to the fact that it does not include in its formulation the kinetic  pressure of the plasma column, fails to make the link between the dynamics and electric behaviour of the plasma column with some thermodynamical aspects involved in the development of the discharge. In other words,  no thermodynamics information results from solving the snowplow equations. Of course, some attemps to compute the temperature and pressure of the plasma column on the basis of the dynamics and electric data obtained from solving the snowplow equations have been done, however those computations are not straightforward really. 

In this work, we consider first the problem of devising an algorithm to predict the temperature of the plasma in experiments on discharges in Z pinch devices. Our solution to that problem consists in to add at the right place of the snowplow equations the missed term that accounts for the kinetic pressure of the plasma. The inclusion of that term yields a set of modified snowplow equations that are suitable to fully describe the physics that goes on in discharges performed through Z pinch devices. This is, by means of the so modified snowplow equations we can straightforwardly compute not only dynamics and electric functions related with the evolution of the plasma column but we can also access related thermodynamics information.

The second issue we treat in this work concerns the possibiity to find out a scaling law for the temperature of the plasma in Z pinch discharges. This is, we analyze the way in which the energetic and spatial wingspan of a Z pinch experimental setup relate with the
maximum temperature a plasma may reach in the associated experiment. We found out that the mere enlargement of the spatial and electrical specifications of the Z pinch experiment do not yield a higher temperature of the plasma. This result is after all consistent with the status that thermodynamics attributes to temperature. This is, temperature is regarded as an intensive thermodynamic quantity. 

In Section 2, we summarize the essential issues on the pinch effect and we examine also the way in which that phenomenon manifests in the context of axial discharges. The section continues with the presentation of a detailed derivation of the snowplow equations for Z pinch systems. In this line, we provide also the derivation of the modified snowplow equations. In closing the section, we confront the data obtained numerically from the modified snowplow equations with the data gathered from the relevant experiments. In section 3, we analyze general Z pinch discharges in the scheme of the modified snowplow equations and we then report on the main results of that analysis. In section 4, we summarize the conclusions this work yields. 

\section{Theory and models}
\subsection{The pinch effect}
Magnetic fields result from the motion of charged particles. In effect, a particle of charge $q$ that moves with velocity $\textbf{v}$ will produce a magnetic field at the point $\textbf{r}$ given by 
\begin{equation}
\textbf{B}(\textbf{r})=\frac{\mu_0q\textbf{v}\times\textbf{r}}{4\pi|\textbf{r}|^3}
\end{equation} 
where $\textbf{r}$ is the vector that goes from the charged particle to the point where $\textbf{B}$ is measured.
A second particle of charge $Q$ and velocity $\textbf{u}$ moving through the region where the first particle has set up the magnetic field $\textbf{B}$ undergoes a force given by
\begin{equation}
\textbf{F}=Q\textbf{u}\times\textbf{B}.
\end{equation}
This law predicts, for instance, that two charges of the same sign moving parallel to each other will undergo an attractive force. By extension, two parallel filaments of current, each of them being considered as a collection of moving charges, also tend to attract to each other. The phenomenon manifests also in the case where there exists not just two parallell filaments of current but there are many of them. In such a case, every single filament of current attracts to all the rest of them, in a way that the self constriction of the bundle of filaments of current takes place. This phenomenon is known as the pinch effect \cite{jackson}.

\subsection{Axial discharges}
A linear pinch experiment also known as a Z pinch experiment is carried out by means of a setup consisting of a cylindrical vessel, whose axis coincides with the $z$ axis of the cylindrical coordinates system, filled with a light gas of interest. This gas is then, by means of a device external to the vessel, subject to an intense electric field which is turned on along the $z$ direction.
As a consequence, the gas ionizes and form a plasma whose positive charges (the ions) will move in one sense along the $z$ direction and whose negative charges (the electrons)
will move in the opposite sense along the $z$ direction. In this way,  each ion and each electron will create its own  magnetic field whose common feature is that anyone of them points in the azimuthal counterclockwise direction around its path in the $z$ direction. In this manner, the magnetic field created by any individual particle will couple with all the other particles of the plasma exerting on each of them, independently of whether the charge of those particles is positive or negative, a force of attraction because the ions are positive charges moving in the positive $z$ direction whereas the electrons are negative charges moving in the negative $z$ direction.   

More specifically, a practical Z pinch device or linear pinch apparatus consists of a cylindrical vessel, or discharge tube, made typically of insulating materials like pyrex, silica, porcelain, etc. At both ends of that discharge tube, there is an electrode made of copper or brass. One of these electrodes is, in addition, connected to a metallic cylinder which closely surrounds the discharge tube, so that it serves as a return conductor for the dicharges triggered within the discharge tube \cite{glasstone}. 

The discharge tube is then filled with a gas of interest which might be either very light as is in the case of deuterium or heavier as is for instance the cases of xenon, argon, helium and others.
The experiment consists in dicharging, through a spark gap, a bank of capacitors of voltage $V_0$ connected to the electrodes (see Figure 1).
\\\\
\begin{figure}[!h]
\includegraphics[width=\textwidth]{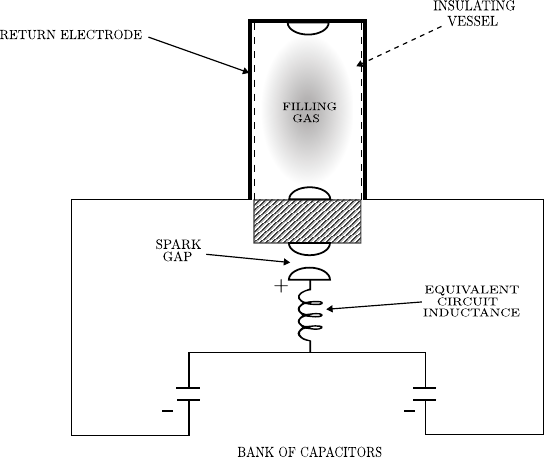}
\caption{Schematic of the experimental setup}
\label{FIG1}
\end{figure}

\pagebreak
When the voltage of the bank of capacitors is high enough, the gas inside the discharge tube ionizes rapidly and converts into a plasma of neglectable resistance. Thereafter, the axial electric field set by the bank of capacitors forces the plasma to flow in a particular fashion. Indeed, owing to the fact that the plasma is a nearly perfect conductor the skin effect comes into play, so that the flow of plasma takes place only within a very thin cylindrical layer adjacent to the inside wall  of the discharge tube. 
The azimuthal magnetic field resulting from that flow of current is nonzero only on the surface of the current sheath but it is equal to zero elsewhere in the plasma. Thus, the azimuthal magnetic field and the current sheath couple together with the result that a force of magnetic origin establishes on the current sheath. More precisely, the coupling between the current sheath and its own magnetic field results in a net force inward that acts on the current sheath as a whole and tends to push it towards the axis of the tube.   

\subsection{The snowplow model for Z pinch systems} 

In general terms, the snowplow model is an analytical scheme devised for describing the dynamics of discharges throughout  confined plasmas. In what follows, we introduce the snowplow model as applied to axial discharges.

To begin with, we consider a cylindrical discharge vessel whose axis coincides with the z axis of the cylindrical coordinates system. The inside radius of this vessel is $r_0$, its outside radius is $R$ and its length is $l_0$. Furthermore, this discharge vessel contains the gas of interest, for the particular experiment we are dealing with, at room temperature $T_0$ and filling pressure $P_0$. As is shown in Figure 1, at each end of the vessel there is an electrode in direct contact with the gas. One of these electrodes is in turn connected to a metallic cylindrical tube of inside radius $R$ that closely surrounds the discharge vessel. In order to produce a discharge through the gas, an intense electric field in the z direction is applied. This is achieved in practice by connecting the electrodes, through a spark gap, to the bank of capacitors charged at the high voltage $V_0$. In this way, the system resembles a coaxial line where the plasma plays the role of the central cable whereas the metallic cylinder surrounding the discharge vessel serves as the return cable.

For the purpose of simplifying the present discussion, it is convenient to assume that the gas inside the vessel is since the beginning fully ionized and therefore its corresponding resistance is neglectable. Under this premise, the application of the high voltage $V_0$ to the plasma leads to the immediate formation of an infinitely thin sheath of plasma carrying a total current $I$ resulting from electrons moving in one sense along the z direction and ions moving in the opposite sense along the z direction as well. This infinitely thin sheath of current is, owing to the skin effect, initially located adjacent to the inside wall of the vessel and therefore its radius is $r_0$. The magnetic field $B_{\theta}$ induced by the current $I$ points in the $\theta$ direction -\textit{i.e.} it is an azimuthal field- and its magnitude is given by
\begin{equation}
B_{\theta}=\frac{\mu_0I}{2\pi r_0}.
\end{equation}
In turn, this magnetic field yields a magnetic pressure $p_M$ that pushes the current sheath inward and whose magnitude is computed through the expression
\begin{equation}
p_M=\frac{B_{\theta}^2}{2\mu_0}.
\end{equation}  
This magnetic pressure forces both the electrons and the ions composing the current layer to move towards the axis of the discharge vessel, in a way that the radius of the current layer $r(t)$ becomes a function of time $t$ and the magnitude of the current $I(t)$  itself becomes also a function of time.

According with the second law of dynamics, the derivative of the momentum of the current layer, regarded as a lump, respect to time must equals the instantaneous force acting on it. This is
\begin{equation}
\frac{d}{dt}\left(M(t)\frac{dr}{dt}\right)=F(t)
\end{equation}
where $M(t)$ represents the time dependent mass of the current layer whereas $F$ corresponds to the instantaneous magnetic force acting on it. 
Furthermore, in the scheme of the snowplow model \cite{rosenbluth}, it is assumed that as the cylindrical current layer moves inward, it behaves as a magnetic piston which sweeps up all the charged particles it encounters. In this manner, the total mass piled up in the current layer, when its radius has contracted fom the initial value $r_0$ to the value $r$, is given by
\begin{equation}
M(t)=\rho_0\pi(r_0^2-r^2)l_0
\end{equation}
where $\rho_0$ stands for the value of the mass density of the filling gas.

On the other hand, the instantaneous magnetic force inward on the current layer is 
\begin{equation}
F(t)=-\left(\frac{\mu_0 I^2 l_0}{4\pi r}\right)
\end{equation}
therefore, implementation of Equation (5) yields the expression
\begin{equation}
\frac{d}{dt}\left(\rho_0\pi (r_0^2-r^2)l_0 \frac{dr}{dt}  \right)=-\left(\frac{\mu_0 I^2 l_0}{4\pi r}\right).
\end{equation}

It is apparent that this single equation is not  enough to find out $r(t)$ since that function is coupled with the other unknown function of the problem, namely the time dependent intensity of  current $I(t)$.

To really solve the problem, one has to supplement a second equation relating $r(t)$ with $I(t)$. Such an equation results from examination of the circuital features of the system. 
Indeed, the problem at hand considers a bank of capacitors of capacity $C_0$ and initial voltage $V_0$ which
is discharged through the gas enclosed in the cylindrical container or discharge tube. 
As we already mentioned, the plasma enclosed within the cylinder is assumed to have no resistivity, however it has inductance which, for the Z-pinch apparatuses, has the form characteristic of a coaxial line, namely 
\begin{equation}
L_p(t)=\frac{\mu_0 l_0}{2\pi}\ln{\left(\frac{R}{r}\right)}.
\end{equation}
In addition, the cabling that connect the bank of capacitors with the tube as well as the intrinsic inductance of the bank of capacitors itself contribute all togheter with  a constant term $L_0$ to the total inductance of the system $L(t)$. This is 
\begin{equation}
L(t)=L_0+L_p(t). 
\end{equation}
Thus, the electrical configuration of the system under consideration is equivalent to that of a capacitor connected in series with a time dependent inductance, so that the instantaneous voltage in the capacitor must equal the voltage in the inductance. This is,

\begin{equation}
V_0-\frac{1}{C_0}\int_0^t I(t')dt' =\frac{d(\Phi(t))}{dt}
\end{equation}
where $\Phi(t)$ stands for the instantaneous magnetic flux through the inductance, namely
\begin{equation}
\Phi(t)=L(t)I(t).
\end{equation}

To make the snowplow equations amenable for computations, it is convenient to cast them in a dimensionless fashion.
To do it, we choose to measure time in units of
\begin{equation} 
t_0=\sqrt{L_0C_0},
\end{equation} 
to measure current in units of 
\begin{equation}
I_0=V_0\sqrt{\frac{C_0}{L_0}}
\end{equation}
and to measure length in units of $r_0$.

After introducing these prescriptions, the snowplow equations can be worked out to set them in a dimensionless form. We did in that way resulting the following set of dimensionless snowplow equations:
\begin{equation}
\frac{d^2r}{dt^2}=\frac{2r^2 \left(dr/dt\right)^2-\alpha^2 I^2}{r(1-r^2)}
\end{equation}
and
\begin{equation}
\frac{dI}{dt}=\frac{1-\int_0^tI(t')dt'+\beta (I/r)(dr/dt)}{1-\beta\ln{(r)}}
\end{equation}
where we have assumed that the thickness of the discharge vessel is negligible, \textit{i.e.} $R\approx r_0$, so that we used $r_0$ instead of $R$ as the radius of the return path for the current. 

The dimensionless parameters $\alpha$ and $\beta$ that appear on the right hand side of the equations above are given, in terms of the physical specifications of the system, by the expressions
\begin{equation}
\alpha=\sqrt{\frac{\mu_0 C_0^2  V_0^2}{4\pi^2  r_0^4\rho_0 }}
\end{equation}
and
\begin{equation}
\beta=\left(\frac{\mu_0 l_0}{2\pi L_0}\right).
\end{equation}
At this stage, it is worth to mention that the normalization of the variables $r$, $I$ and $t$ is not affected by changes of either electrical or geometrical specifications of the experiment under consideration. Any those changes manifests only through the dimensionless parameters $\alpha$ and $\beta$ which in turn modulate the snowplow equations,  so that such changes may indeed 
yield different yet normalized $r(t)$ and $I(t)$ functions.
  
The numerical solution for the dimensionless functions $r(t)$ and $I(t)$ can be readily obtained by supplementing the snowplow equations with the following initial conditions
\begin{equation}
r(0)=1,
\end{equation}
\begin{equation}
I(0)=0,
\end{equation}
\begin{equation}
\left(\frac{dr}{dt}\right)_{t=0}=0,
\end{equation}
\begin{equation}
\left(\frac{d^2r)}{dt^2}\right)_{t=0}=-\left(\frac{\alpha}{\sqrt{3}}\right)
\end{equation}
and
\begin{equation}
\left(\frac{dI}{dt}\right)_{t=0}=1.
\end{equation}

\subsection{Modified snowplow equations for Z-pinch systems}
Since in the snowplow scheme introduced above the kinetic pressure of the plasma has not been considered, the corresponding snowplow equations predict, at first, the collapse of the current sheath onto axis without any opposition. However, that situation departs from what has been observed experimentally. What has been really observed under suitable experimental conditions is that the current sheath first accelerates and detaches from the wall of the container. Then, as time goes on, it contracts up to reach a radius of the order of $10\,\%$ of its initial radius $r_0$. Following the formation of this first pinch, the current sheath usually bounces several times. This is, the current sheath expands partially and then it contracts again successively until owing to the onset of instabilities breaks up completely. The time elapsed in this bouncing process corresponds somehow to what is considered as the confining time of the plasma \cite{glasstone}.

The explanation for the dynamics described above is as follows, during the contraction of the current sheath the plasma enclosed by it undergoes a process of adiabatic compression which in turn tends to heat up that plasma. Therefore, the kinetic pressure of the plasma rises up resulting in a force outward on the current sheath which in this manner prevents its free collapse onto axis. The successive bounces of the current sheath result from the competition between the magnetic pressure and the kinetic pressure as the plasma column cools down and heats up, alternatively. 

To take into account the role of the kinetic pressure of the plasma in the course of the collapse of the current sheath, we must add on the right hand side of Equation (8) a term that somehow accounts for the force of kinetic origin acting on the current sheath. 
To find out what such a force must be about, we start by recognizing, on the one hand, that
each particle of the plasma that undergoes an inelastic collapse with the current sheath gains a kinetic energy $e_k$ given by
\begin{equation}
e_k=\frac{1}{2}mv^2
\end{equation}
where
\begin{equation}   
v=\frac{dr}{dt}
\end{equation}
stands for the velocity of the current sheath whereas $m$ represents the mass of the ions. We ignore at this instance the electrons because their mass is much smaller than the mass of the ions.

On the other hand, we also assume that the current sheath is completely efficient to pile up all the particles (the ions say) it encounters. Thus, as the current sheath reaches the radius $r$ all the particles of the system get enclosed by it. In such a situation, the average number density of ions $n$ must satisfy the relationship
\begin{equation}
(\pi r^2 l_0)n=(\pi r_0^2 l_0)n_0 
\end{equation}
where $n_0$ the initial number density of ions and therefore the term on the right hand side of Equation (26) corresponds to the the total number of ions $N_0$.
Thus, we obtain the relationship that exists between the average number density of ions and the radius of the current sheath, namely
\begin{equation}
n=\left(\frac{r_0}{r}\right)^2 n_0.
\end{equation}
Hence, when the current sheath moves from position $r$ to position $r-dr$, it piles up
\begin{equation} 
dN(t)=-\left[l_0 2\pi r(t) dr\right] n(t)
\end{equation}
particles which contribute an amount
\begin{equation}
dE_k=e_k dN
\end{equation}
to the kinetic energy of the gas. To compute the kinetic energy accumulated by the plasma during the interval of time $[0,t]$, we have to carry out the integral
\begin{equation}
E_k(t)=\int_0^t\frac{1}{2}mv^2dN
\end{equation}
which can be cast in the fashion
\begin{equation}
E_k(t)=-\pi  r_0^2 l_0\rho_0\int_0^t\frac{1}{r}\left(\frac{dr}{dt'}\right)^3dt'
\end{equation}
where $\rho_0$ is the mass density of the filling gas, so that -by ignoring the mass of the electrons- its value essentially coincides with the value of the mass density of the ions of the filling gas after this has been transformed into a plasma. In this way, the value of $\rho_0$ is simply given by $\rho_0=mn_0$.

At this point, we postulate that a process like conduction heat transfer sets up. This is, the particles provided with more kinetic energy share, by means of successive collisions at random, part of their kinetic energy with the particles provided with less kinetic energy. In this manner, assuming the best scenario in which the motion of the ions becomes really random and the kinetic energy is well distributed among the ions, we can assert that the thermodynamic function internal energy $U$ for this system is numerically equal to the kinetic energy gained by the ions upon compression.

In that situation, since the plasma is a gas it must satisfy the following relationship between its kinetic pressure
\begin{equation}
p_k=\frac{2}{3}\left(\frac{U}{V}\right),
\end{equation}
its volume 
\begin{equation}
V=\pi r^2(t) l_0 
\end{equation}
and the internal energy it gains upon compression
\begin{equation}
 U(t)=-\pi r_0^2l_0\rho_0\int_0^t\frac{1}{r}\left(\frac{dr}{dt'}\right)^3dt'.
\end{equation}
Hereafter, the expression internal energy will refer to the internal energy gained by the gas upon compression only. We will ignore the internal energy the gas had before its compression because that piece of internal energy is very much smaller than $U$.

The force outward on the current sheath $F_k$ is simply given by the product of the kinetic pressure on the current sheath times the area of the current sheath. That product leads to the expression
\begin{equation}
F_k(t)=-\frac{4\pi r_0^2 l_0\rho_0}{3r}\int_0^t\frac{1}{r}\left(\frac{dr}{dt'}\right)^3dt'.
\end{equation}
We have added this force on the right hand side of the snowplow equation for $r(t)$, Equation (8), and then we have worked out it, as well as the equation for $I(t)$, to obtain the dimensionless version for the modified snowplow equations. The result is the  set of equations, namely
\begin{equation}
\frac{d^2r}{dt^2}=\frac{6r^2 (dr/dt)^2-3\alpha^2 I^2-4\int_0^t (1/r)(dr/dt')^3 dt'}{3r(1-r^2)}
\end{equation}
and
\begin{equation}
\frac{dI}{dt}=\frac{1-\int_0^tIdt'+\beta (I/r)(dr/dt)}{1-\beta\ln{(r)}}.
\end{equation}
These equations supplemented with the corresponding initial conditions, mentioned earlier, can be solved numerically to obtain the dimensionless functions $r(t)$ and $I(t)$. Moreover, once we have the dimensionless $r(t)$ and $I(t)$ we can compute the quantity that, for the goal of our study, really matters. This is, the internal energy gained by the gas since the beginning of the discharge and the time when the current sheath reaches it maximum compression. That quantity is given by Equation (34) which in terms of dimensionless $r$ and $t$ transforms into
\begin{equation}
U=-\left(\frac{\beta}{\alpha^2}\right)E_0\int_0^t\frac{1}{r}\left(\frac{dr}{dt'}\right)^3dt'.
\end{equation}

\subsection{Simulations in the framework of the modified snowplow equations versus actual experiments}
The first thing to notice is that the dimensionless functions $r(t)$ and $I(t)$ are enterely determined by the dimensionless equations of either the snowplow model or the modified snowplow model and also by the values attained by the dimensionless parameters $\alpha$ and $\beta$ that appear in those equations. For instance, the question about whether a particular discharge gives rise to the formation of a pinch can be readily answered on the basis of what the values of the parameters $\alpha$ and $\beta$, for that specific experiment, are. Therefore, if one changes some of the physical specifications of a given system, in a sort that the values for $\alpha$ and $\beta$ do not undergo any alteration then the functions $r(t)$ and $I(t)$ will remain necessarily the same as they were before the introduction of those modifications. In regard to the actual physical magnitudes, the value of the time when the system pinches, in $[s]$, the value of the maximum current reached in the discharge, in $[A]$, the value of the internal energy gained by the gas upon compression in, $[eV]$, etc will all change according with both the energetic content and the physical size of the system. This is, the particular values for the quantities just listed will depend on the physical specifications of the system. Nevertheless, characteristic phenomena such as the formation of a pinch and the features of the dynamics evolution of the current sheath will be preserved under modifications of the experimental setup if those modifications leave $\alpha$ and $\beta$ invariants. In other words, different systems that however share the same values of $\alpha$ and $\beta$ behave, as far as their dynamics is concerned, similarly.

In order to test in practice the suitability of the set of modified snowplow equations, we have integrated numerically those equations for the case of a generic experiment whose specifications fall within the range of values that a number of actual experiments, carried out through linear discharge apparatuses, share (see for instance \cite{glasstone}, chapter 7 and references therein). This is, we have considered a cylindrical tube of  about $3\, L$ volume which is filled with helium gas at a pressure of about $1\, mm$ of mercury and at room temperature of $300\, K$. The electrodes, sited at both ends of the cylinder, are connected through a spark gap to a bank of capacitors subject to a difference of potential of about $12\, [ kV]$, in a way that the gas inside the vessel undergoes an electric field that falls within the range of some few hundreds volts per centimeter. The characteristic inductance of the tube plus the cabling is about $30\, [nH]$ and the total capacity of the bank of capacitors is about $80\, [\mu F]$. Since we are mainly interested in the study of the implosive phase of the plasma, we assume for the sake of simplicity that the gas inside the tube is already ionized when the high voltage is applied on it. Thus, the value for the resistance of the gas is considered to be negligible small.

On the basis of these physical specifications, we obtain that the total electrostatic energy stored in the bank of capacitors $E_0$ is about $5760\, [J]$ which corresponds to $3.6\times 10^{22}\, eV$ whereas the total number of particles (\textit{i.e.} helium ions plus electrons) inside the cylinder is of the order of $N=3\times10^{20}$. Hence, even in the rather optimistic but unrealistic situation in which all the electrostatic energy stored in the bank of capacitors $E_0$ could be somehow converted into internal energy of the gas $U$, the temperature of this would be irretrievably bounded. Indeed, the temperature of the gas $T$ and the internal energy of the gas $U$ relate through the equation
\begin{equation}
\frac{3}{2}\left(k_BT\right)=\left(\frac{U}{N}\right)
\end{equation}
therefore, in the hypothetical situation mentioned, the value of  $k_BT$ would be of the order of $80\, eV$ if $N$ comprises ions plus electrons. Otherwise, if $N$ accounts only for the ions then the value of $k_BT$ would shift to $k_BT\approx240\, eV$.
In any case, each particle would have, in average, a kinetic energy of the order of the hundreds of $eV$ which is clearly far away from the figure of hundreds $keV$ per particle which is what is required for setting up a self sustained thermonuclear reaction . 

Of course, the experiments carried out under the physical conditions listed above should be intended not for the goal of attaining through them thermonuclear fusion, let alone to attain through them self sustained thermonuclear fusion, but they should only be regarded as a tool for exploring the subject of thermonuclear fusion and hopefully for getting an insight on it.

By fine tuning the physical specifications listed above, we have found by means of our numerical simulations that a notorious pinch of the plasma column results when the dimensionless parameters $\alpha$ and $\beta$, computed on the basis of those specifications, take the values $\alpha=4.08$ and $\beta=2.23$.

Figure 2 displays the evolution of the radius $r(t)$ of the current sheath with time $t$ whereas Figure 3 shows the magnitude of the current $I(t)$ flowing across the current sheath as a function of time. Both curves reproduce well the data obtained from actual experiments (see \cite{glasstone}, chapter 7). The main point to notice is that the plasma sheath first contracts slowly because the current flowing through it is still small and so is the magnetic pressure acting on it.
As the current flowing through the current sheath increases the magnetic pressure on it also increases and then the current sheath is enforced to contract rapidly. However, during that contraction the gas enclosed by the current sheath gains internal energy that translates into kinetic pressure acting on the current sheath that in turn opposes the magnetic pressure applied on it. In this way, the contraction of the current sheath is slowed down until it gets stopped. After that, the current sheath continues for a while its motion in the form of a sucession of expansions and contractions. In successful experiments as well as in our simulation the extent of the first contraction of the plasma is such that the radius of the current sheath reduces to a tenth approximately with respect to its initial value. This is, the radius of the current sheath at the first contraction is approximately $0.1 r_0$.

At this point, it is worth to stress that the agreement between our simulation on a typical discharge in a Z pinch device and its corresponding experimental counterpart reinforces retrospectively the relevance of the modified snowplow equations. In effect, the modified snowplow equations are based on the expression we have postulated for accounting for the internal energy and the kinetic pressure of the plasma. In turn, the internal energy and the kinetic pressure of the plasma depend on the functions $r(t)$ and $I(t)$ present in the snowplow equations. Hence, the snowplow equations conform an algorithm for determining self consistently the internal energy of the plasma, the kinetic pressure of the plasma, $r(t)$ and $I(t)$. In this connection, the good agreement between the curves $r(t)$ and $I(t)$ obtained from solving the snowplow equations and their corresponding counterparts taken from an actual experiment suggests that the internal energy of the plasma computed through the snowplow equations would also coincide with the internal energy of the plasma in the corresponding actual experiment. Needless to say that at present that comparison is not available because the internal energy of the plasma in actual experiments is not measurable yet. 
In any case, for the purpose of the discussion that follows, we adopt the criterion that the internal energy computed self consistently through the modified snowplow equations fits the internal energy of actual experiments on discharges in Z pinch devices.
\pagebreak
\begin{figure}[!h]
\includegraphics[width=.82\textwidth]{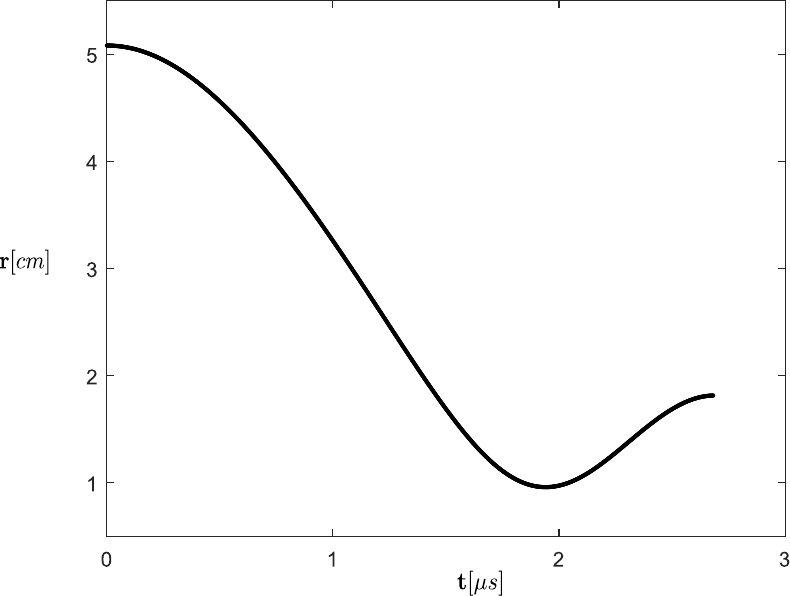}
\caption{Radius of the current sheath versus time.}
\label{FIG2}
\end{figure}
\begin{figure}[!h]
\includegraphics[width=.90\textwidth]{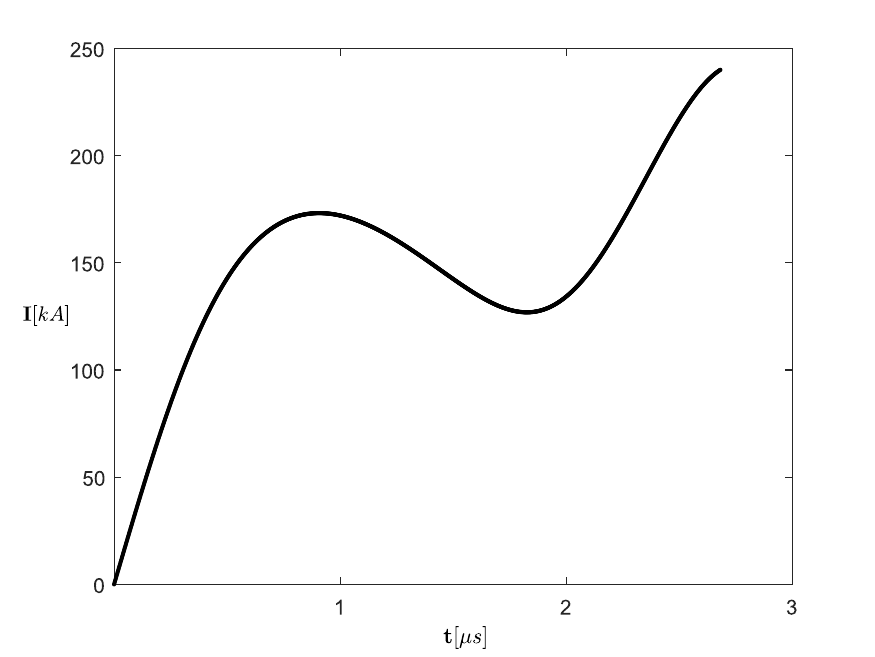}
\caption{Current flowing across the current sheath across the time.}
\label{FIG3}
\end{figure}

\pagebreak
\section{Results and Discussion}
After having established the suitability of the modified snowplow equations for computing the curves $r(t)$ and $I(t)$, we are in a position to go ahead by using those functions to evaluate the various energy terms that play a role in determining the physical behaviour of the systems under study. 

To start with, we may regard the Z pinch system as composed of two pieces, namely a bank of capacitors of mostly capacitive impedance $Z_C$ and a load, that comprises the discharge tube, the cabling and the plasma, of basically inductive impedance $Z_L$. As the bank of capacitors is discharged through its load, a current flows in the plasma and a flux of power sets up in the system. Of course, the rate at which the bank of capacitors delivers power to its load is governed by the rules of power transfer between a source of impedance $Z_C$ and load of impedance $Z_L$.

The instantaneous power $P_d(t)$ delivered by the bank of capacitors to its load is given by
 \begin{equation}
P_d(t)=\frac{d}{dt}\left[L(t)I(t)\right]I(t)
\end{equation}
therefore the energy delivered during the time interval $[0,t]$ results from evaluating the integral
\begin{equation}
E_d(t)=\int_0^t P_ddt'.
\end{equation}
This energy can in turn be decomposed into two pieces, namely $E_s$ and $E_i$ where $E_s$ corresponds to the energy that gets stored in the inductance of the system and it is computed as
\begin{equation}
E_s(t)=\frac{1}{2}\left[L_0+L_p(t)\right]I^2
\end{equation}
whereas $E_i$ which is obtained as
\begin{equation}
E_i(t)=E_d(t)-E_s(t)
\end{equation}
 is that really contributes to the dynamics of the plasma. 

However, it is not all the energy $E_i$ that contributes to the internal energy of the plasma but on the contrary, it is only a fraction of $E_i$ that enters the plasma in the form of internal energy. This fact, technically speaking, reflects through the specific evolution that $r$ and consequently $n$ have along the duration of the discharge. The remaining fraction of $E_i$ is spent in accelerating the current sheath inward.

For the purpose of having equations amenable for computations, we cast  the expressions for the energies mentioned above in the following fashion:
\begin{equation}
E_d(t)=-2\beta E_0\int_0^t\frac{1}{r}\left(\frac{dr}{dt'}\right)I^2dt'+2E_0\int_0^t\left(1-\beta\ln{(r)}\right)I\left(\frac{dI}{dt'}\right)dt',
\end{equation}
\begin{equation}
E_i(t)=-\beta E_0 \int_0^t\frac{1}{r}\left(\frac{dr}{dt'}\right)I^2dt'
\end{equation}
where 
\begin{equation}
E_0=\frac{1}{2}C_0V_0^2
\end{equation}
corresponds to the energy stored initially in the bank of capacitors whereas the various integrands that appear on the right hand side comprise only dimensionless variables. Let us now call the internal energy gained by the plasma during the lapse of time $[0,t]$ as $U_0$, therefore the corresponding rise of temperature is obtained as
\begin{equation}
k_BT_0=\frac{2}{3}\left(\frac{U_0}{N_0}\right).
\end{equation}
Hence, according to Equation (38), the practical expression to compute the temperature of the plasma reads
\begin{equation}
k_BT_0=-\frac{2}{3}\left(\frac{\beta}{\alpha^2}\right)\left(\frac{E_0}{N_0}\right)
\int_0^t\frac{1}{r}\left(\frac{dr}{dt'}\right)^3dt'.
\end{equation}
These formulas and in particular the last one for calculating the temperature of the plasma allow us to analyze the feasibility that a particular system attains thermonuclear fusion. Indeed, as it has already been mentioned the figure of merit for the purpose that a plasma develops a process of thermonuclear fusion is the actual magnitude of its temperature whose computation can be carried out by means of Equation (48).
In order to appreciate the wealth of possibilities provided by this formalism, let us consider the following situation:
\\ Suppose we know all the electrical and geometrical specifications of a totally optimized Z pinch experiment. Therefore, we can compute the values of the dimensionless parameters $\alpha$ and $\beta$ straightforwardly and then we can write down the specific modified snowplow equations appropriate to model that experiment. After solving numerically those equations for the functions $r(t)$ and $I(t)$, we can compute $k_BT$ at the time of the pinch which, in passing, is a goal \textit{per se} because so far there is no definitive technique for measuring the temperature of the plasma in Z pinch discharges.

Earlier in this article, we applied this protocol to the case of a typical Z pinch experiment in helium in which the bank of capacitors was charged at a voltage $V_0\approx 12\, kV$ storing an electrostatic energy $E_0\approx 3.6\times 10^{22}\,eV$. In that occasion, we reported only the results we obtained for $r(t)$ and $I(t)$. Now, we complete that report. 

We have used $r(t)$ and $I(t)$ to compute the internal energy that the gas gained upon compression. What we obtained was that only a minor fraction of the electrostatic energy stored initially in the bank of capacitors was converted into internal energy of the gas. Specifically, in the lapse of time that goes from the beginning of the discharge to the time of maximum contraction of the current sheath, the internal energy of the gas grew up to reach the value $U\approx 0.14E_0$. In regard to the temperature of the gas at the time when the current sheath reached its maximum contraction, the result was $k_BT_0\approx 30\,eV.$  Notice that to compute this temperature we have divided the internal energy gained by the ions by the number of ions. We have disregarded the presence of electrons that in any case would contribute, by means of the mechanism of bremsstrahlung radiation, to cool the gas down. 

The result for the temperature of the gas at pinching, namely $k_BT_0\approx 30\,eV$, is relatively significant. In effect, in the hypothetical case in which all the electrostatic energy stored in the bank of capacitors transforms into internal energy, the temperature of the gas would reach the value $k_BT\approx 240\, eV$. This is, $k_BT_0=0.125\, k_BT$.

It is worth to stress once again that self sustained thermonuclear fusion of a mixture of deuterium and tritium requires as a necessary but not sufficient condition that the plasma temperature reaches a value of at least $20\,keV$ whereas self sustained thermonuclear fusion of plasmas made up of elements other than hydrogen isotopes requires necessarily temperatures much in excess higher than that. Hence, the experiment just mentioned is far away to meet the minimal condition for attaining thermonuclear fusion. Nevertheless, one might well hypothesize that by means of 
an enlarged experimental setup, the temperature and  density of the plasma could rise up, in such a way as to eventually configurate the conditions for triggering a proccess of thermonuclear fusion. 

Of course, there exist several ways of enlarging an experimental setup, however here for the sake of simplicity we limit ourselves to consider only a particular family among those possible scalations. We proceed in the following manner:
we select a reference experiment from the class of fully optimized experiments.  For instance, the prototypical experiment discussed above could be perfectly eligible as the reference experiment. Then, on the basis of that choice, we consider an scaled experiment subject to the couple of constraints, namely\\
1) the discharge tube of the scaled experiment is filled with the same gas at the same density $\rho_0$ (\textit{i.e.} same number density $n_0$) as in the case of the reference experiment\\
and\\
2) the changes of the physical size and energy content of the scaled experiment are such that the values of $\alpha$ and $\beta$ are correspondingly the same as those of the reference experiment.

Under these constraints, both the reference experiment and the scaled experiment develop the same dynamics. Now,
the obvious way in which these constraints can be fulfilled under scaling up is by increasing by the same factor, $\lambda>1$ say, all the physical specifications of the reference experiment with the exception of  its filling density $\rho_0$. 

In this manner, the physical specifications of the scaled up experiment transform as: 
\begin{equation}
C_0\to\lambda C_0,
\end{equation}
\begin{equation}
 V_0\to\lambda V_0,
\end{equation}
\begin{equation}
 L_0\to\lambda L_0,
\end{equation}
\begin{equation}
 r_0\to\lambda r_0
\end{equation}
 and 
\begin{equation}
l_0\to\lambda l_0.
\end{equation}
Therefore, the energy stored in the bank of capacitors passes from the value
\begin{equation}
E_0=\frac{1}{2}C_0V_0^2
\end{equation} 
in the reference experiment to the value
\begin{equation}
E=\lambda^3E_0
\end{equation}
in the scaled up experiment.
In a similar fashion, the number of plasma particles within the discharge tube changes from the value
\begin{equation}
N_0=\pi r_0^2 l_0 n_0
\end{equation}
in the reference experiment to the value
\begin{equation}
N=\lambda^3 N_0
\end{equation}
in the scaled up experiment.

On the other side, in accord with the modified snowplow model, the solutions for $r(t)$ and $I(t)$ in the reference experiment must coincide exactly with their corresponding counterparts in the scaled up experiment. Hence, the internal energy $U_0(t)$ gained by the plasma in the reference experiment relates with the internal energy $U(t)$ gained by the plasma in the scaled up experiment through the equation
\begin{equation}
U(t)=\lambda^3U_0(t).
\end{equation}
Therefore, providing that the value of $\lambda$ is sufficiently large, the gas in the scaled up experiment may gain an enormous amount of internal energy. However, the number of particles of the enlarged experiment increases also as $\lambda^3$ times the number of particles of the reference experiment, then no matter what the value of $\lambda$ is, the temperature of the gas in the scaled up experiment, given by
\begin{equation}
k_BT=\frac{2}{3}\left(\frac{U}{N}\right)
\end{equation}
 reduces to the expression for the temperature of the gas in the reference experiment, namely
\begin{equation}
k_BT_0=\frac{2}{3}\left(\frac{U_0}{N_0}\right).
\end{equation}
In synthesis, according with the modified snowplow model, no change in temperature results from this way of scaling up a Z pinch experiment.

This somehow disappointing result motivates us to look for an alternative way of thinking about an enlarged experiment. 

Once again, we start from a given reference experiment and we practice then the following modifications on some of its specifications:

\begin{equation}
C_0\to\lambda_1 C_0,
\end{equation}
\begin{equation}
V_0\to\lambda_1 V_0,
\end{equation}
\begin{equation}
r_0\to\lambda_1 r_0
\end{equation}
where $\lambda_1>1$ and the filling gas as well as its density is the same as that of the reference experiment, so that the value of $\alpha$ is preserved.

In order to preserve also the value of the dimensionless parameter $\beta$, let us consider the following modifications:
\begin{equation}
l_0\to\lambda_2 l_0
\end{equation}
and
\begin{equation}
L_0\to\lambda_2 L_0
\end{equation}
where $\lambda_2>1$.

Of course, the feasibility to scale the total inductance of the apparatus up may not be guaranteed \textit{a priori} but on the contrary it could be subject to all sort of practical restrictions. On the other side, when $\lambda_1\neq \lambda_2$ the resulting scaling up resembles a change in the shape of the enlarged experiment as compared with the shape of the reference experiment.
This matter is in some way or another controversial because we have assumed from the beginning that the reference experiment we have chosen was already fully optimized. This is, both the electrical specifications of the reference experiment and the shape of it were carefully chosen to obtain from the resulting experimental setup the best possible performance.  

In any case, if this scaling up through $\lambda_1$ and $\lambda_2$ is indeed factible, then the energy stored in the bank of capacitors changes from the value
\begin{equation}
E_0=\frac{1}{2}C_0V_0^2
\end{equation}
in the reference experiment to the value
\begin{equation}
E=\lambda_1^3 E_0
\end{equation}
in the scaled up experiment whereas the number of particles in the reference experiment
\begin{equation}
N_0=\pi r_0^2 l_0 n_0
\end{equation}
grows as
\begin{equation}
N=\lambda_1^2 \lambda_2 N_0
\end{equation}
in the enlarged experiment. 

Thus, the performance of the enlarged experiment as compared with that of the reference experiment  yields more internal energy. The specific relationship between those internal energies is 
\begin{equation}
U=\lambda_1^3U_0.
\end{equation}
In this manner, the relationship between the temperature attained by the reference experiment and that attained by the enlarged experiment becomes
\begin{equation}
k_BT=\left(\frac{\lambda_1}{\lambda_2}\right)k_BT_0.
\end{equation}
To have a glimpse on how the extent of the electrostatic energy stored in the bank of capacitors influences the temperature reached by the plasma in Z pinch experiments, we work out  Equation (67) in combination with Equation (71) to obtain the equation
\begin{equation}
\left(\frac{k_BT}{k_BT_0}\right)=\frac{1}{\lambda_2}\left(\frac{E}{E_0}\right)^{\frac{1}{3}}.
\end{equation}

Once again, this result is not that auspicious as far as the goal of devising an apparatus able to achieve thermonuclear fusion is concerned. Indeed, if the performance of an optimized actual apparatus leads to temperatures of the order of $k_BT_0\approx 10\,eV$ say, then it would be desirable to scale that apparatus up in the hope it could perform in such a way that leads to temperatures of the order of $k_BT\approx 10\, keV$. However, according with the scaling law summarized by Equation (72), a rise of the temperature of the plasma by a factor $10^3$ would require an increase of the energy stored in the bank of capacitors by a factor $10^9$ which, in actual practice, is prohibitive. 

In order to unravel what the ingredients that determine the temperature achieved by the gas in a Z-pinch discharge are, we notice in the first place that according with the principle of conservation of energy the inequality
\begin{equation}
U_0<E_0
\end{equation}
holds.
Here, we recall that $U_0$ is the internal energy gained by the gas during the time elapsed between the beginning of the discharge and the time  when the first pinch happens approximately whereas $E_0$ is the energy strored initially in the bank of capacitors.

In regard the scheme of the modified snowplow model, complete swept up of the particles of the gas is assumed, so that the temperature of the gas at pinching can be computed as
\begin{equation}
k_BT_0=\frac{2}{3}\left(\frac{U_0}{N_0}\right)
\end{equation}
and therefore the constraint 
\begin{equation}
k_BT_0<\frac{2}{3}\left(\frac{E_0}{N_0}\right)
\end{equation}
follows.

Next, by using Equation (48) for computing the internal energy, we derive the following inequality
\begin{equation}
\left(\frac{\beta}{\alpha^2}\right)\left(-\int_0^t \frac{1}{r}\left(\frac{dr}{dt'}\right)^3\right) dt')<1.
\end{equation}
In appearance, the factor that restricts the temperature a Z pinch discharge may attain is the ratio $(E_0/N_0)$ mainly. However, in an interrelated manner other factors also play necessarily a role. Indeed, the electric and spatial specifications of the given system under analysis, which are accounted altogether by $\alpha$ and $\beta$, are the ingredients that through the modified snowplow equations determine the dynamics of the discharge and consequently they determine also the temperature of the plasma when the pinch happens.

In closing, let us stress that on the basis of the specifications collected from many optimized experiments carried out along decades, the big picture might be depicted as follows:

\begin{equation}
\frac{2}{3}\left(\frac{E_0}{N_0}\right)\sim 10^2\, eV
\end{equation}
and
\begin{equation}
\left(\frac{\beta}{\alpha^2}\right)\sim 1.
\end{equation}
The corresponding simulations within the modified snowplow model yields to
\begin{equation}
-\int_0^t\frac{1}{r}\left(\frac{dr}{dt'}\right)^3dt'\sim 1,
\end{equation}
so that the expected temperature that could be achieved through those experiments would be $k_BT_0\sim 10^2\,eV$ only.

\section{Conclusions}

To clearly set the limits of validity of our results, we stress that in the present work we adopted the snowplow model for studying experiments on discharges in Z pinch devices. Below, we list the main results we have obtained.

\begin{enumerate}
\item We adapted the snowplow equations in a way to obtain from their solutions not only dynamics and electric information on the  plasma but also thermodynamics information about it
\item We verified that the solutions of those modified snowplow equations reproduces well the dynamics and electrical data gathered from actual Z pinch experiments
\item We estimated that the temperature a plasma may reach through Z pinch experiments is of the order of dozens of $eV$ only
\item We proved that the enlargement by a common factor of the spatial and electrical parameters of a Z pinch device does not modify the value of the temperature that a plasma may eventually reach in the associated experiment
\item By considering a particular way of enlarging a refererence Z pinch experiment, which in passing it would involve a change in the shape of the corresponding Z pinch device, we obtain a scaling law for the temperature that the plasma may reach. That scaling law indicates that that temperature scales with the cubic root of the energy stored initially in the bank of capacitors that feeds the discharge. In this manner, the idea that thermonuclear fusion is reachable by means of experiments performed through Z pinch devices, regardless its size and its energetic content, should be discarded.
\end{enumerate}

\end{document}